# PLASMA KINETICS IN THE ETHANOL/WATER/AIR MIXTURE IN "TORNADO" TYPE ELECTRICAL DISCHARGE

D Levko[1], A Shchedrin[1], V Chernyak[2], S Olszewski[2], O Nedybaliuk[2]

[1]Institute of Physics National Ukrainian Academy of Sciences, 03028, Kiev, Ukraine

[2]Taras Shevchenko National University, 03122, Kiev, Ukraine

*E-mail:* d.levko@gmail.com

**Abstract**

This paper presents the results of a theoretical and experimental study of plasma-assisted reforming of ethanol into molecular hydrogen in a new modification of the "tornado" type electrical discharge. Numerical modeling clarifies the nature of the non-thermal conversion and explains the kinetic mechanism of nonequilibrium plasma-chemical transformations in the gas-liquid system and the evolution of hydrogen during the reforming as a function of discharge parameters and ethanol-to-water ratio in the mixture. We also propose a scheme of chemical reactions for plasma kinetics description. It is shown that some characteristics of the investigated reactor are at least not inferior to characteristics of other plasma chemical reactors.

## 1. Introduction

The interest to alternative fuels research in the last two decades is increased by the depletion of the traditional fossil fuels. Today, ethanol is considered the most perspective fuel for internal-combustion engines [1]. First, it could be produced from renewable sources (biomass, industrial waste, etc). Second, its combustion produces relatively small amount of pollutants. However, the low velocity of the ethanol combustion wave propagation does not allow its use in its pure form as an engine fuel [2]. In order to increase this velocity, one needs to enrich $C_2H_5OH$ by molecular hydrogen [3], since the latter has higher flame speed than alcohol. Unfortunately, there is another problem of storing $H_2$ on a vehicle. Recently, several methods were proposed to obtain hydrogen from hydrocarbon fuels before its entering to the engine [1]. The methods are partial oxidation, steam reforming, dry $CO_2$ reforming, thermal decomposition and plasma-assisted reforming.

The use of non-equilibrium plasma looks more attractive as a result of its lower energy consumption. Plasma acts as a catalyst and initiates fast chain reactions that do not progress under normal conditions. Today different plasma chemical reactors are used (for example [1], [4-6]) for molecular hydrogen generation from different hydrocarbons (ethanol, methane, etc) in non-equilibrium plasma. In [7-8], a new plasma chemical reactor for ethanol-to-hydrogen conversion was proposed. It was shown that such reactor had high energy efficiency. Unfortunately, electrical discharge was unstable with high currents. That instability leads to the fast electrode destruction and equipment fail.

This work presents the results of experimental and theoretical investigation of plasma kinetics in new plasma chemical reactor. This device uses the reverse vortex gas flow of "tornado" type similar to Fridman's group [9]. Discharge of such type operates in the transitional thermal-to-nonthermal plasma regime. It is characterized by a presence of electrons with average electron energy of few electron volts and a neutral gas temperature of ≈1000-2000 K [10]. High gas temperature increases hydrogen production as a result of the additional conversion of hydrocarbons generated in the discharge region. The presence of active O, OH and H in the mixture converts efficiently the hydrocarbons in the post-discharge region. The use of vortex air flow increases the stability of the discharge.

An advantage of the proposed reactor over the one at [9, 10] is the use of a special working chamber. This chamber allows conversion of hydrocarbons either in the liquid or in the gas phase. In case of liquid, part of the input power goes into evaporation. The presence of liquid isolates the metallic electrode from the plasma region. It also prevents the electrode erosion and increases the working time of the reactor. Also, the reactor does not require the additional gas heating in the pyrolysis camera, because the post-discharge region with the high gas temperature acts as this camera.

**2. Experiments**

Figure 1 shows the scheme of the experimental setup. It consists of a cylindrical quartz vessel *(1)* with a diameter of 9 cm and height of 5 cm, sealed by flanges at the top *(2)* and at the bottom *(3)*. The vessel is filled with working liquid *(4)* (ethanol/water mixtures at different ratios) through the inlet pipe *(5)*. The level of liquid is controlled by a spray pump. The basic cylindrical T-shaped stainless steel electrode *(6)* is fully immersed into the liquid and is additionally cooled by water. The electrode on the upper flange *(2)* is made from duralumin and has a special copper hub *(11)* with the axial nozzle *(7)* which is of diameter 2 mm and length of 6 mm.

The air is injected into the vessel through the orifice *(8)* in the upper flange *(2)* tangentially to the cylinder wall *(1)* and creates a reverse vortex flow of tornado type. Rotating gas *(9)* descends to the liquid surface and moves to the central axis where it flows out through the nozzle *(7)* in the form of jet *(10)* into the quartz chamber *(12)*. Since the area of minimal static pressure above the liquid surface during the vortex gas flow is located near the central axis, it creates a column of liquid at the gas-liquid interface in a form of a cone with its height of ~1 cm above the liquid surface (without electric discharge).

The voltage is applied between the upper electrode *(2)* and the lower electrode *(6)* in the liquid with the help of the DC power source powered up to 10 kV. Only one mode of the discharge operation is studied. It is the mode with "solid" cathode (SC): "-" is on the flange *(2)*. The conditions of breakdown in the discharge chamber are regulated by three parameters: the level of the work liquid, the gas flow rate $G$, and by the value of voltage $U$. The ignition of discharge usually begins at the appearance of the axial streamer. Transition time to the self-sustained mode of operation is ~1-2 s. The range of discharge currents varied within 100-400 mA. The pressure in the discharge chamber during the discharge operation is ~1.2 atm, static pressure outside the reactor is ~1 atm. The elongated ~5 cm plasma torch *(10)* is formed during the discharge burning in the camera which does not contain any oxygen.

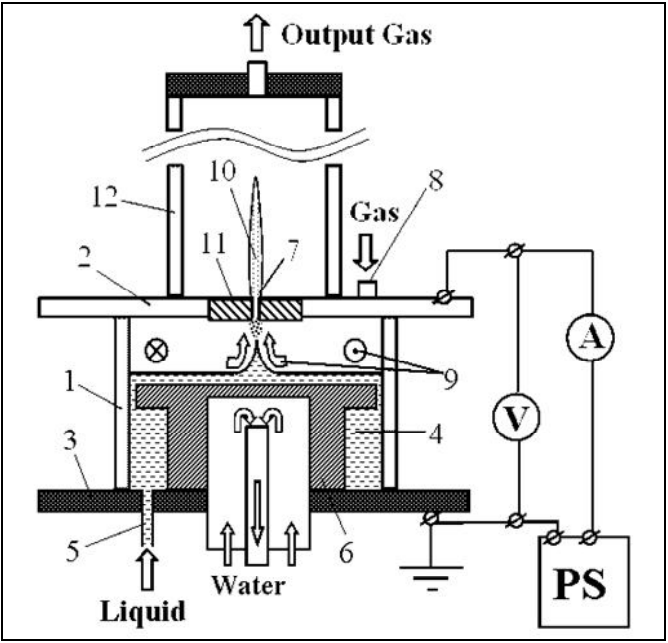

**Figure 1.** Scheme of the experimental setup

Figure 2 shows the typical current-voltage characteristics of the "tornado" type electrical discharge in the mode with "solid" cathode, working in water at different airflow rates. Notice that the dependence is typical for glow discharge [11].

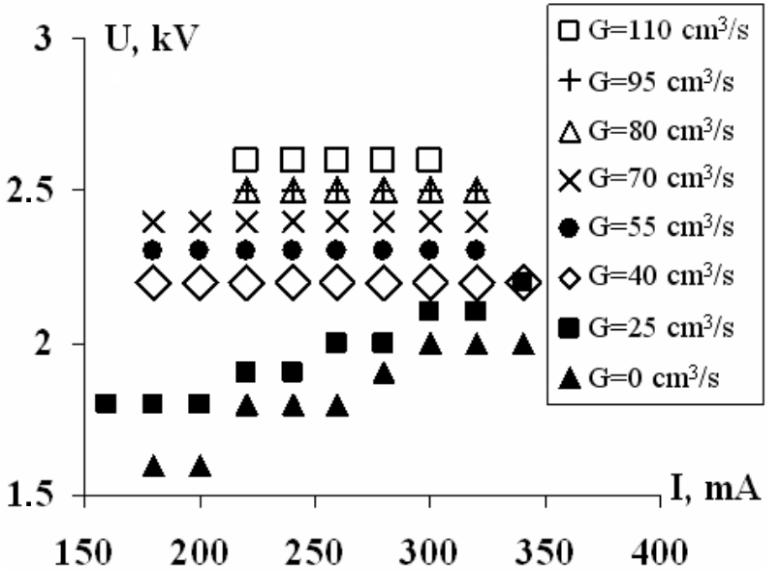

**Figure 2.** Current-voltage characteristics in the regime with solid cathode working in water at different airflow rates

Experimentally it is obtained that for 6.5%, 13% and 26% concentration of ethanol in liquid the breakdown conditions do not change significantly. For $G$ = 55 cm$^3$/s the breakdown voltage is 2 kV and the current is 320 mA. But the ethanol flow rate changes. It is equal to 1 ml/min for 6.5% and 2 ml/min for 13% and 26%.

**3. Model of calculations**

We use a global model in our calculations. We assume that the discharge is homogeneous over the entire volume. It is justified at zero approximation, because the time of the gas mixing in radial direction is less than the times of characteristic chemical reactions. Also we neglect the processes in the transitive zone between the discharge to post-discharge. The volume of the transitional zone is much smaller than the volume of discharge and post-discharge zones. Thus, the time of gas pumping through the transition region is too short for the chemical reactions to have a sufficient influence on the concentration of neutral components.

The total time of calculation is divided into two time intervals: the first one is the calculation of the kinetic processes of fast generation of active atoms and radicals in the discharge region. Those components accelerate the formation of molecular hydrogen, carbon oxides and other hydrocarbons production. The second time interval is the oxidation of the gas mixture in the post-discharge region as a result of the high gas temperature and the presence of O and OH. These components remain in the mixture after the dissociation of water and oxygen molecules by electrons impacts in plasma.

The oxidation of generated hydrocarbons (mainly $C_xH_y$ and formaldehyde $CH_2O$) has noticeable influence on kinetics in the investigated mixture due to high gas temperature (~1000 K). Under aforementioned conditions, the characteristic time of oxidation is approximately equal to the air pumping time through the discharge region (~10$^{-3}$-10$^{-2}$ s). Therefore, the low temperature plasma model [7, 8], where the continuous discharge is divided into the sequence of quasiconstant discharges, is not applicable here. The following system of kinetic equations is used in order to account the constant air pumping through the system:

$$\frac{dN_i}{dt} = S_{ei} + \sum_j k_{ij} N_j + \sum_{j,l} k_{ijl} N_j N_l + ... + K_i - \frac{G}{V} N_i - kN_i. \qquad (1)$$

It is calculated using a solver developed at the Institute of Physics NAS of Ukraine. That solver was verified many times on other systems and has demonstrated good results. $N_i$, $N_j$, $N_l$ in eq. (1) are concentrations of molecules and radicals; $k_{ij}$, $k_{iml}$ are rate constants of the processes for $i$-th component. The rates of electron-molecule reactions are

$$S_{ei} = \frac{W}{V} \frac{1}{\varepsilon_{ei}} \frac{W_{ei}}{\sum_i W_{ei} + \sum_i W_i}, \qquad (2)$$

where $W$ is the discharge power and $V$ is the discharge volume. In this model, $W$ is the full power $W_0$ divided by factor of two. This division corresponds to the most stable regime of the discharge [11]. The value of $W_0$ is determined by the current-voltage characteristic (figure 2). $W_{ei}$ is the specific power deposited into the inelastic electron-molecular process with threshold energy $\varepsilon_{ei}$:

$$W_{ei} = \sqrt{\frac{2q}{m}} n_e N_i \varepsilon_{ei} \int_0^\infty \varepsilon Q_{ei}(\varepsilon) f(\varepsilon) d\varepsilon . \qquad (3)$$

Here $q = 1.602 \cdot 10^{-12}$ erg/eV, $m$ is the mass of electron and $n_e$ is the concentration of electrons. The variable $Q_{ei}$ is inelastic process cross section, $f(\varepsilon)$ is the electron energy distribution function (EEDF); $W_i$ is the specific power deposited into elastic processes:

$$W_i = \frac{2m}{M_i} \sqrt{\frac{2q}{m}} n_e N_i \int_0^\infty \varepsilon^2 Q_i(\varepsilon) f(\varepsilon) d\varepsilon . \qquad (4)$$

Here $M_i$ are the molecules' masses, $Q_i$ are the transport cross sections for nitrogen, oxygen, water and ethanol molecules.

The last three terms in eq. (1) describe the constant inflow and outflow of gas from the discharge region. The term $K_i$ is the inflow of molecules of the primary components (nitrogen, oxygen, water and ethanol) into the plasma, $G/V \cdot N_i$ and $kN_i$ are the gas outflow as the result of the air pumping and the pressure difference between the discharge region and the atmosphere.

In order to define the initial conditions, the ethanol/water solution is assumed to be an ideal solution. Therefore, the vapours' concentrations are linear functions of the ethanol-to-water ratio in the liquid. The evaporation rates $K_i$ of $C_2H_5OH$ and $H_2O$ are calculated from the measured liquids consumption. The inflow rates $K_i$ of nitrogen and oxygen are calculated by the rate of air pumping through the discharge region:

$$K_i = \frac{G}{V} N_i^0, \qquad (5)$$

where $N_i^0$ correspond to $[N_2]$ and $[O_2]$ in the atmospheric pressure air flow.

In the non-equilibrium plasma almost the entire energy is deposited into the electron component. The active species, generated in the electron-molecular processes, lead to chain reactions with ethanol molecules. The released energy heats the gas. Calculations show that the fastest reactions of that type are

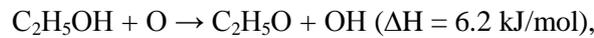
$C_2H_5OH + O \rightarrow C_2H_5O + OH$ ($\Delta H = 6.2$ kJ/mol),

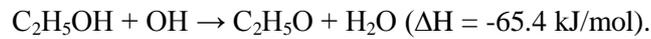
$C_2H_5OH + OH \rightarrow C_2H_5O + H_2O$ ($\Delta H = -65.4$ kJ/mol).

The first reaction is endothermic and the second is exothermic. By comparing their specific enthalpies, one can estimate the average gas temperature, which is near 1200 K. The gas temperature in the discharge region is taken to be constant in the model. In reality the gas temperature $T$ is dependent on the gas pumping rate and the heat exchange with the environment. Therefore, in order to account of those influences, $T$ is varied in the interval 800-1500 K (similarly to experimentally obtained temperature spread).

After $\sim 10^{-2}$ s the balance between the generation and the decomposition of the components leads to saturation of concentrations of all species. It allows to stop the calculations in the discharge region and to investigate the kinetics in the post-discharge region. The system (1) is solved without accounting of the last three terms on the time interval without plasma. The calculations are terminated when the molecular oxygen concentration reaches zero level. That time interval equals to a few milliseconds. The gas temperature in this region is found using the equation [2, 12]:

$$\frac{dT}{dt} = -\frac{1}{\rho C_p} \sum_i H_i(T) \cdot \mu_i \frac{dN_i}{dt}. \tag{6}$$

Here $\rho$ is the gas density, $C_p$ is the average specific gas heat capacity under constant pressure, $H_i$ and $?_i$ are molar enthalpy and molar mass of *i*-th component respectively.

It is seen from eq. (3)-(4) that the specific powers $W_{ei}$ and $W_i$ are influenced by the electron energy distribution function. EEDF is calculated from Boltzmann kinetic equation in the two-term approximation for breakdown field in the air [13]. The last assumption was made due to the fact that nitrogen is a plasma-forming gas. Only the processes with the primary components (see table 1) are taken into account in EEDF calculations. The cross sections of processes 17-19 are absent in the recent literature. Therefore, in order to approximate these cross sections, we used the approximation [13]. At this approximation the cross section of ethanol is equal to the cross section of oxygen biased on the doubled threshold energy.

Figure 3 shows the calculated EEDFs for different gas temperatures and 6.5% ethanol concentration in solution. It is shown that EEDF is defined by vibrations of water molecules (two levels ((100)+(010)) and (010)), excited by electrons, whose energy is less than 0.05 eV. In the 1.5-2 eV region EEDF is defined by the excitation of electron-vibrational levels of nitrogen and water by electron impact. When the electrons' energy is higher than 8 eV EEDF looks like Maxwellian distribution function. At this region the function's behavior is defined by the dissociation and ionization of molecules by electrons' impacts. The average electrons' energy is 0.55 eV for 800 K, 0.46 eV for 1200 K and 0.43 eV for 1500 K.

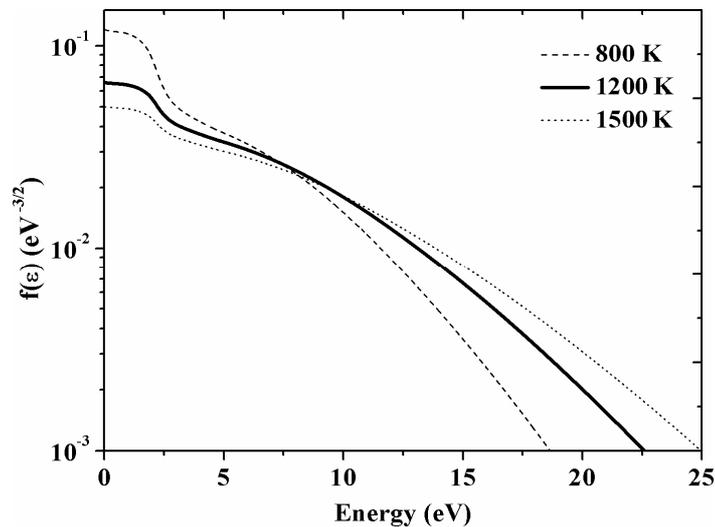

**Figure 3.** Calculated electron energy distribution functions at 6.5% ethanol concentration in the liquid

There is no generally accepted kinetic mechanism for modeling plasma kinetics in air/water/ethanol mixture. In our previous works [7-8], [22], we proposed the mechanism [23] for low temperature region. However, this mechanism does not describe kinetics in "tornado" type electrical discharge, where the gas temperature is much higher. Marinov's mechanism [24] and Dagaout's ethanol sub-mechanism [25] could be more appropriate for describing the chemical reactions in the investigated mixture. Our preliminary calculations have shown that these schemes include many unneeded components which do not contribute to the end product in the investigated regimes. They may be ignored in the scheme of reactions. In [26], Dunphy proposed a kinetic mechanism for high-temperature ethanol oxidation. The investigated temperature interval (1080-1660 K) in that article is close to the one at present work. Therefore, Dunphy's scheme was chosen as the basic mechanism in our research. It was expanded by Held's methanol sub-mechanism [27] and by additional important processes in hydrogen oxidation [28]: $OH+OH+M \rightarrow H_2O_2+M$, $OH+OH \rightarrow H_2O+O$, $H+H+M \rightarrow H_2+M$, $H_2+O_2 \rightarrow OH+OH$, $H_2+O_2 \rightarrow H+HO_2$. The full mechanism developed in this work is presented in table 2. It is composed of 30 components and 130 chemical reactions between them. The charged particles (electrons and ions) were ignored in the mechanism, because of low degree of ionization of the gas ($\sim 10^{-6}$-$10^{-5}$). Nitrogen acts as the third body in the recombination and thermal dissociation reactions. Additionally, the nitrogen-containing species were removed from the mechanism, since they are not the subject of this study. Also, our previous research [46] has shown that the processes between these components and hydrocarbons were third-order reactions.

It is well known that ethanol dissociates into three isomeric radical of $C_2H_5O$ ($CH_3CHOH$, $CH_2CH_2OH$ and $CH_3CH_2O$), with comparable probabilities. However, in our research only one radical ($CH_3CHOH$) was chosen and denoted as $C_2H_5O$. The rate constants of the chemical reactions with this isomer have the highest values.

## 4. Results and discussion

The process of ethanol pyrolysis is described by the following formula:

$$C_2H_5OH + 1/2 O_2 \rightarrow 2CO + 3H_2.$$

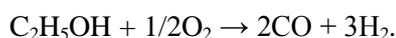

In this research the mixture is far from that stoichiometry. When ethanol concentrations in the liquid are 6.5%, 13% and 26%, then ethanol-to-oxygen ratios in gas phase are equal to $\sim$ 0.1, $\sim$ 0.2 and $\sim$ 0.5, respectively. These ratios are the ratios between the initial ethanol and oxygen concentrations in the discharge region which are defined by the rates of primary components pumping through the discharge (5). Namely, the mixture does not satisfy the stoichiometry and there is a fuel-poor mixture in discharge region. At the exit of the plasma region, the gas contains oxygen, ethanol, methane and other hydrocarbons. At the post-discharge region the ratio $[C_2H_5OH]/[O_2]$ remains non-stoichiometric, but part of oxygen is used for hydrocarbons pyrolysis.

For the investigated temperature interval, the rate constant $k_1$ of the ethanol thermal dissociation (R1.2, table 2) changes in the range between $10^{-26}$-$10^{-17}$ cm$^3$/s. Third body concentration [M] is near ~$10^{18}$ cm$^{-3}$. The product between [M] and $k_1$ is much smaller than $S_{ei}$ for ethanol dissociation by electrons impacts (see table 3). Consequently, the process of ethanol dissociation is much more efficient than thermal dissociation. As a result, the main channels for active species generation are electron-molecular dissociation of the primary components.

Table 3 shows the list of electron-molecular reactions which were taken into account in the solution of equation (1). Unfortunately, current literature lacks information about cross sections of reactions which are marked by asterisks. To evaluate the rates of such processes we used the technique described in [13]. Threshold energies are twice the energies of broken bonds [48].

Independently from the ethanol-to-water ratio in the solution for the first 10-100 ?s, the molecular hydrogen is mainly generated in the process:

$$C_2H_5OH + H \rightarrow C_2H_5O + H_2. \qquad (6)$$

Table 3 shows that the initial H production goes through two parallel channels: ethanol (R2.3) and water (R27.3) dissociation by electron impacts. The values of $W_{ei}$ and $S_{ei}$ were calculated for ethanol concentration of 13% in the solution. Table 3 shows that the threshold energy of (R27.3) is less than (R2.3), and $S_{ei}$ for the latter is greater than for the former. However, for the investigated ethanol-to-water ratios, $S_{ei}$ factor is overpowered by the ratio between the initial water and ethanol vapours concentrations in the plasma. Additionally, [H$_2$O] increases and [C$_2$H$_5$OH] decreases during the discharge. Therefore, the presence of water in the solution indirectly stimulates the molecular hydrogen generation from ethanol by H atoms.

When $C_xH_y$ concentrations reach high levels, $H_2$ is generated in the H-abstraction processes, which are typical for high-temperature oxidation:

$$C_xH_y + H \rightarrow C_xH_{y-1} + H_2. \qquad (7)$$

Namely, the non-equilibrium plasma stimulates high-temperature processes in a relatively low temperature interval. This stimulation increases the hydrogen yield and removes excess hydrocarbons from the mixture. When the ethanol concentration in the solution is equal to 6.5%, $T > 1200$ K and the time is higher than few milliseconds, then $H_2$ kinetics is defined by the water abstraction process:

$$H_2O + H \rightarrow OH + H_2. \qquad (8)$$

This process takes place due to the increase over time of water concentration in the discharge region. It leads to the growth of the rate of (8) versus the decrease of the rate of (7). So, at low ethanol concentrations in the solution it serves as the gas heater and the initial $H_2$ generator. In the saturation region (figure 4a) the hydrogen is mainly produced from water.

It can be seen at table 2, that the rate constant for the ethylene abstraction has the highest value among the reactions described by (6)-(7). However, calculations have shown that this process has sufficient influence only at temperature of 1200 K. At other temperatures, the main reactions are the abstraction of CH$_4$ and C$_2$H$_6$ and formaldehyde thermal dissociation (R51.2).

Extremums of some components' concentrations (figure 4a) are caused by the reactions between molecules and radicals, and between electrons and molecules. The rates of these processes grow continuously with the increasing concentrations. When they reach a certain level, degeneration exceeds formation and concentrations decline rapidly. This decline stops when the production exceeds decay again. Starting at $t \approx 10^{-2}$ s these processes are in equilibrium and all concentrations reach constant values.

Figure 4b shows, that in the combustion zone the concentrations of the remaining ethanol and oxygen after the discharge continue to decrease, as well as methane concentration. The last occurs as a result of the interaction between $CH_4$ and active species H, O and OH (R46.2)-(R48.2). At the same time, we see a weak growth of $H_2$, $H_2O$ and $CO_2$ concentrations. The value of $[CO_2]$ increases in the carbon oxide oxidation processes (R78.2) and (R79.2).

Figure 4 shows that for some time interval in the discharge region the molecular hydrogen concentration reaches its maximum and then decreases. It is described by oxidation of $H_2$ in the processes (R83.2) and (R84.2), with active OH and O respectively. However, such behavior is not universal at different $T$ values. Figure 5 shows the calculated dynamics of $[H_2]$ for 13% ethanol in the solution at 800 K, 1200 K and 1500 K. It is seen that $[H_2]$ grows continuously when $T = 800$ K. The rate constants of both (R83.2) and (R84.2) reactions change substantially when the temperature is increased from 800 K to 1200 K. It decreases the influence of oxidation processes.

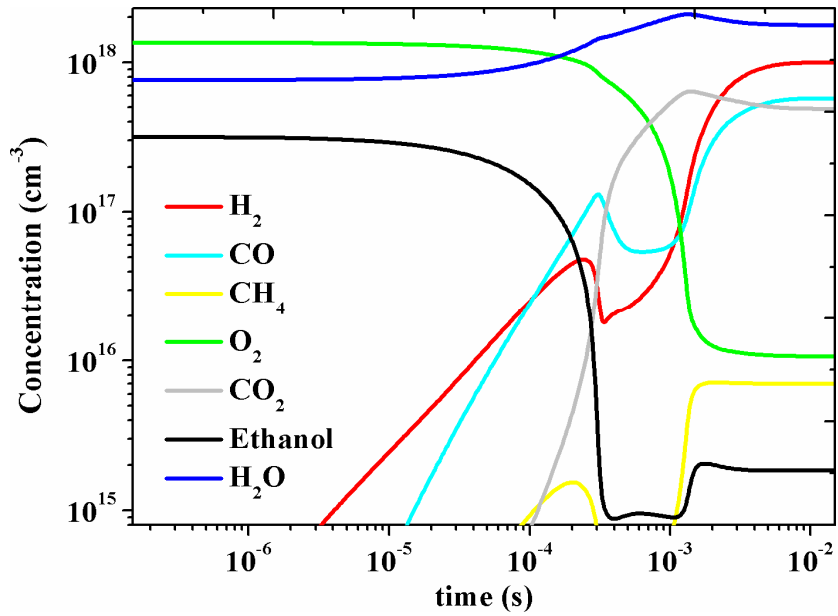

(a)

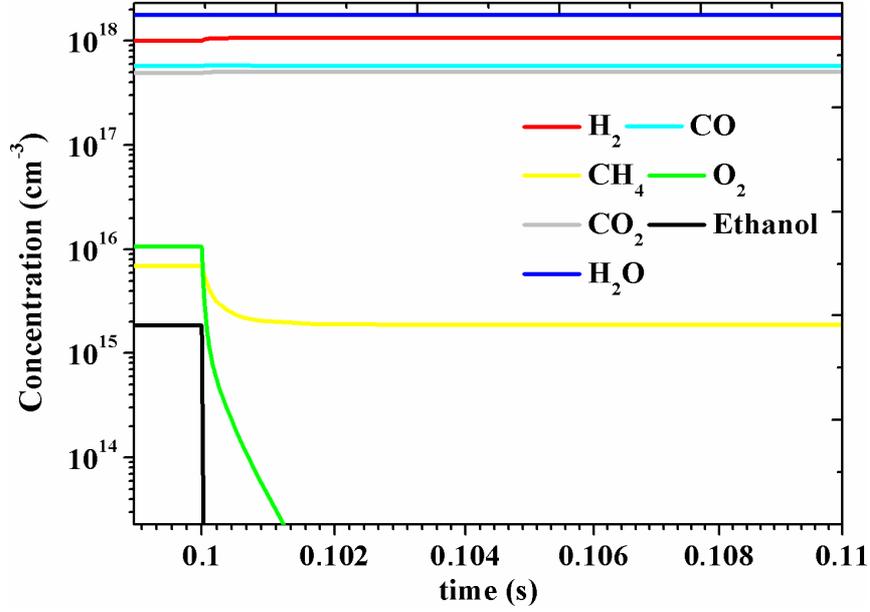

(b)

**Figure 4.** Calculated time evolutions of some stable species concentrations in the discharge (a) and post-discharge (b) regions at $T = 1200$ K and 13% concentration of ethanol in the solution

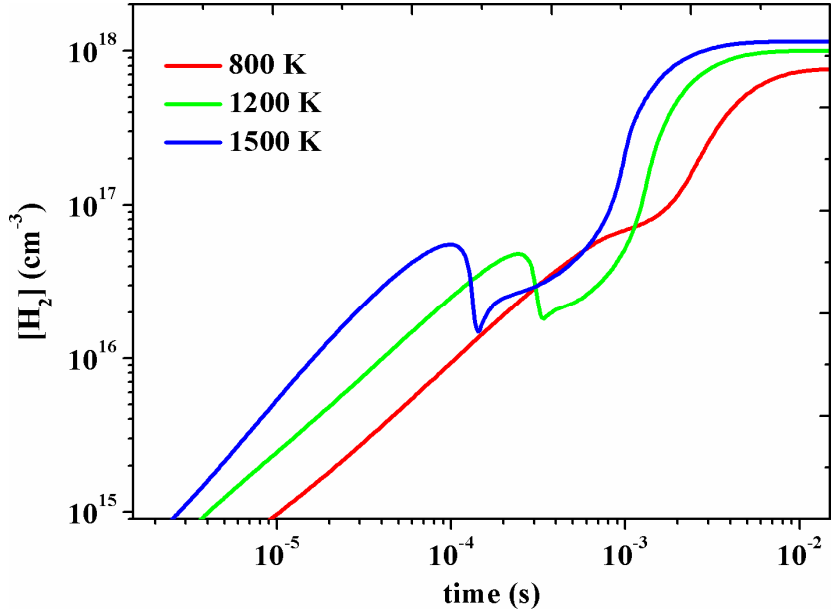

**Figure 5.** Time dependences of hydrogen concentration in the discharge at different gas temperatures

In the plasma region at the first 10-100 ?s the main sources of atoms and radicals (O, OH, H, $CH_3$, $CH_2OH$, $C_2H_5O$, and $C_2H_5$) are electron-molecular dissociation reactions of primary components. Afterwards, the reactions of the radicals thermal dissociation become predominant. These reactions are negligible in the low-temperature plasma ($k \sim 10^{-20}$-$10^{-30}$ cm$^3$/s), but play a key role in the researched "tornado" type electrical discharge ($k \sim 10^{-13}$-$10^{-17}$ cm$^3$/s). For temperatures higher than 1200 K, these processes have much more influence on radicals degeneration than dissociation by electron impacts. Let us consider reactions of $C_2H_5O$ dissociation (R25.2) and (R9.3). The ratio between the rates of these processes are equal to $k(R25.2)/k(R9.3)\cdot\alpha^{-1}$, where α is the ionization degree (it is $10^{-6}$-$10^{-5}$ in the investigated discharge). For temperature increase from 800 K to 1500 K the ratio is in the range from $5\cdot10^{-9}\cdot\alpha^{-1}$ to $3\cdot10^{-6}\cdot\alpha^{-1}$. Namely, thermal dissociation becomes the main channel of $C_2H_5O$ degeneration for some values of $T$.

When the time is higher than 1 ms the chaining of the reactions is realized by the interaction between atoms and radicals, and the stable species generated in the discharge. So, under the investigated conditions the non-equilibrium plasma chemistry is combined with the combustion chemistry. Plasma initiates the chain mechanism and the combustion leads to its development.

For temperatures lower than 1000 K in the investigated mixtures, hydrogen atoms are produced mainly in the process of water dissociation by electron impact:

$$H_2O + e \rightarrow OH + H + e. \qquad (9)$$

At the highest temperatures this process is the prevailing channel only for few microseconds. Afterwards, H is formed from the radicals $C_2H_5O$ and $C_2H_5$. They are generated directly from ethanol by collision with electrons.

Figure 6 shows the calculated time evolution of some active species concentrations. It is seen that their behavior is typical for oxidation. The first active components to appear in the mixture are O, OH and H. The latter two are generated from ethanol and water in processes (R1.3), (R2.3) and (R27.3). Oxygen atoms are generated in reaction (R4.3), when OH radicals are dissociated by electron impacts. Inactive $HO_2$ radicals are not formed directly from the primary species. They are produced by recombination of $O_2$ with hydrogen atoms (R28.2), which cause chain termination. Later, it reacts with some radicals in (R19.2)-(R24.2) and also causes the decrease of concentrations of the active components.

When the mixture enters to the combustion zone, the active species concentrations decrease rapidly (figure 6b), which is caused by the recombination of radicals in three-body reactions. Later, the radicals O and OH oxidize the hydrocarbons and carbon oxide CO. These active components leave the gas mixture faster than H and $CH_3$. Despite the fact that hydrogen atoms take part in the abstraction processes, their concentration decrease slower than [O] and [OH], because they are formed in the thermal dissociation of $C_2H_5$ (R100.2) and $C_2H_3$ (R105.2). Figure 6b shows that the slowest decrease is observed for $CH_3$ radicals. In the post-discharge region they are generated in methane H-abstraction.

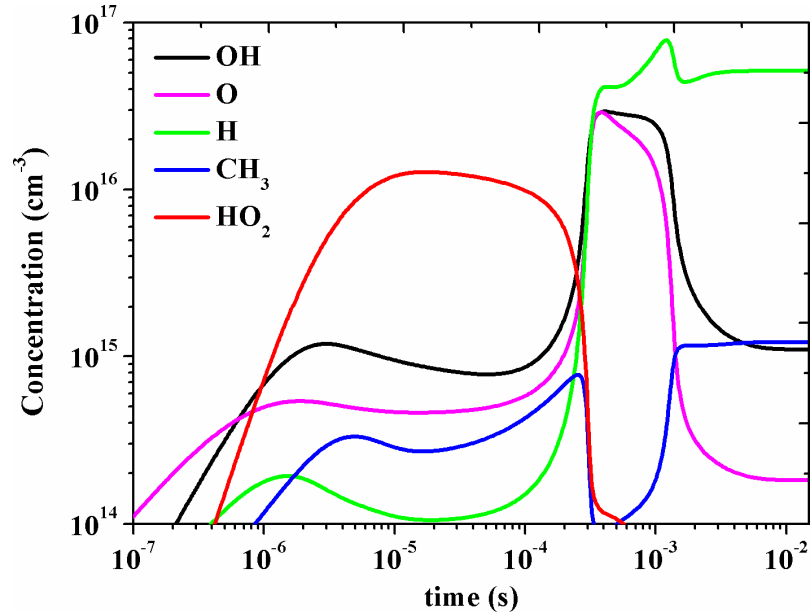

(a)

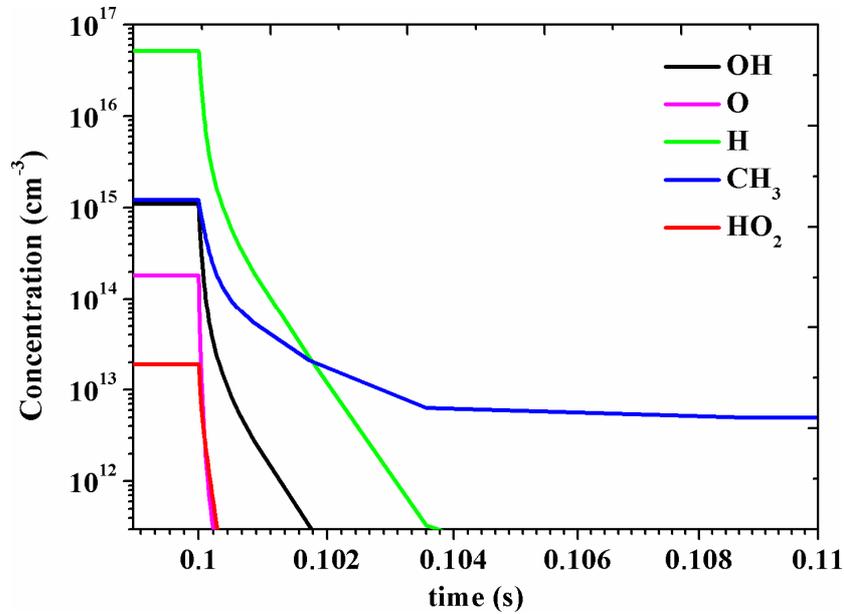

(b)

**Figure 6.** Calculated time dynamics of some active species concentrations in the discharge (a) and post-discharge (b) regions at $T = 1200$ K and 13% concentration of ethanol in the solution

The composition of the gas at the output of the reactor is analyzed by the gas-phase chromatography using a gas chromatograph 6890 N Agilent with calibrated thermal conductivity detectors. The measurements show, that the main components of the mixture that leave the reactor are $H_2$, CO, $CO_2$, $CH_4$, $C_2H_4$, $C_2H_6$ and $C_2H_2$. This result is in good agreement with the results of numerical simulation (figure 4). Figure 7 shows the comparison between experimental and computational data for different ethanol-to-water ratios in the solution. It is seen that the measured data are in a good agreement with computed ones. When ethanol concentration is 6.5% and 13% at $T = 1500$ K, then hydrogen and carbon monoxide are the main components of the gas mixture. At figure 7 we present the sum of concentrations of $H_2$ and CO, since CO converts fully into $H_2$ by water gas shift reaction (CO+$H_2O$ → $H_2$+$CO_2$, [1]). This process was not taken into account in the kinetic mechanism (table 2), because its rate constant is too low to have sufficient influence on kinetics in discharge and post-discharge regions of our reactor. The best agreement between measured and calculated data at 26% ethanol concentration is reached at $T = 1000$ K.

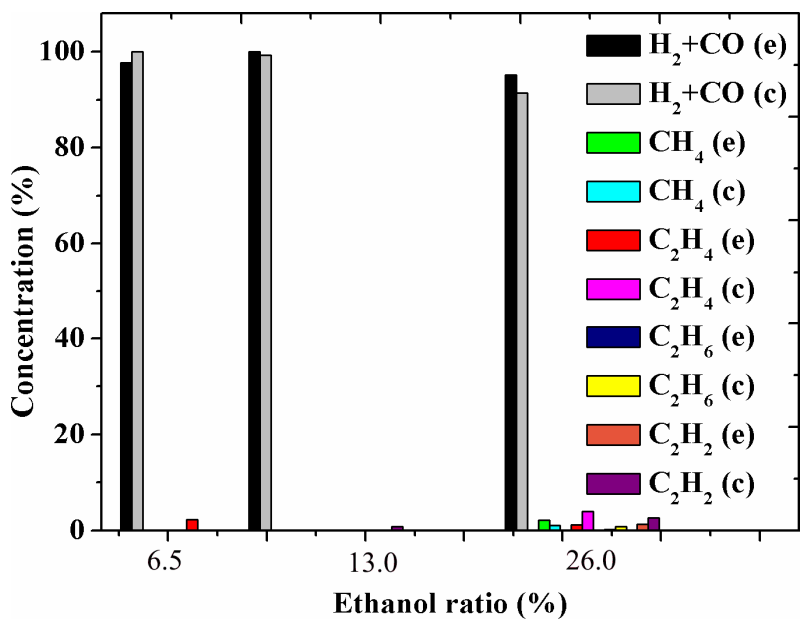

**Figure 7.** Comparison between calculated (c) and experimentally (e) obtained concentrations of main components

In order to define molecular hydrogen yield, the following expression is used [1]:

$$Y(H_2) = \frac{[H_2]}{3 \cdot [C_2H_5OH] + [H_2O]}. \quad (10)$$

Here $H_2$ concentration is taken on the outlet of the reactor, and the values of ethanol and water concentrations are taken at the beginning of the discharge phase. Figure 8 shows the calculated dependences of hydrogen yield from gas temperature for different ethanol concentrations. It is seen that in all three cases the yield $Y(H_2)$ grows when $T$ increases. Additionally, the highest yield is reached when ethanol concentration in the solution is 13% and $T = 1500$ K. Under these conditions $H_2$ generation process is efficient through the abstraction of hydrocarbons (7) and through the water H-abstraction (8).

Figure 9 presents the increase of $H_2$ yield at the post-discharge region as compared to the yield at the discharge region. It is seen that the use of the combustion zone is more beneficial in case of low ethanol concentration in liquid at high temperatures. On one hand, the increase of the ethanol-to-water ratio increases ethanol's flow rate (from 1 ml/min to 2 ml/min) and the initial ethanol vapour concentration. On the other hand, the growth of $[C_2H_5OH]$ leads to a decrease of both $[H_2O]$ and the rate of the channel (9). Figure 9 shows, that the factor of water concentration decrease is much more dominant than the factor of ethanol concentration increase. Therefore, the highest molecular hydrogen yield is reached for 13% concentration of ethanol. These results are in good agreement with the results of figure 8.

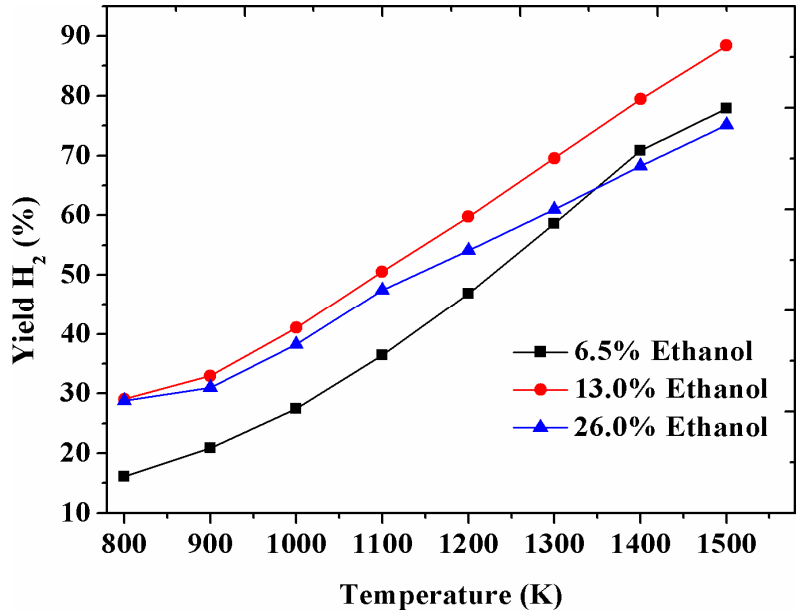

**Figure 8.** The calculated dependence of the molecular hydrogen yield from the gas temperature

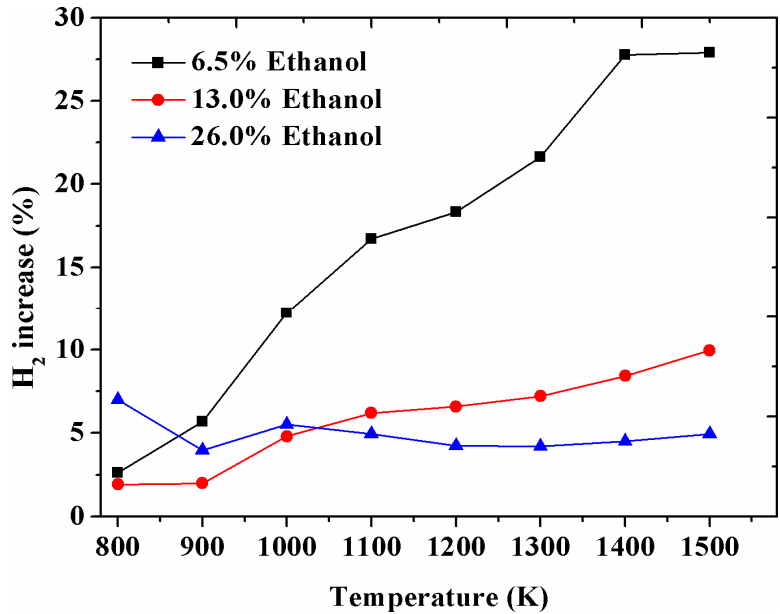

**Figure 9.** The increase of yield of the molecular hydrogen as a result of the use post-discharge region

An important characteristic of plasma chemical reactor is its efficiency [1]:

$$\eta = \frac{(Y_{H2} + Y_{CO}) \times LHV(H_2)}{IPE + Y_{HC} \times LHV(HC)} \quad . \tag{11}$$

Here we take the sum of hydrogen and carbon oxide yields due to water gas shift reaction. Also in (11) *LHV* is the lower heating value of the hydrogen and the fuel (ethanol), *IPE* is the introduced plasma energy. It is seen from figure 10 that the highest $\eta$ value is 6.5% at ethanol concentration of 26%. It is much lower than the values obtained by the Massachusetts Institute of Technology and the Waseda University groups for pure ethanol [1]. However, this reactor has a higher value of conversion rate (see figure 11) than the named groups' reactors. This characteristic is calculated from the expression [1]:

$$\chi = \frac{[CO + CO_2 + CH_4 + ...]_{produced}}{2 \cdot [C_2H_5OH]_{injected}}. \qquad (12)$$

The last formula describes the efficiency of breaking C-C and C-H bonds in ethanol molecule. Figure 11 shows that the highest $\chi$ is reached at 6.5% of [$C_2H_5OH$] in the solution. Starting at 1000 K $\chi$ increases when the temperature grows. Such behavior is associated with the post-discharge region, where ethanol is fully oxidized. The reactions between the stable hydrocarbons and the active species do not change the sum in the numerator of (12). They lead to the redistribution of carbon atoms between carbon oxides and $C_xH_y$. Let us note that $\chi$ value, which can be larger than 100%, is attributed to the constant ethanol pumping through the discharge zone.

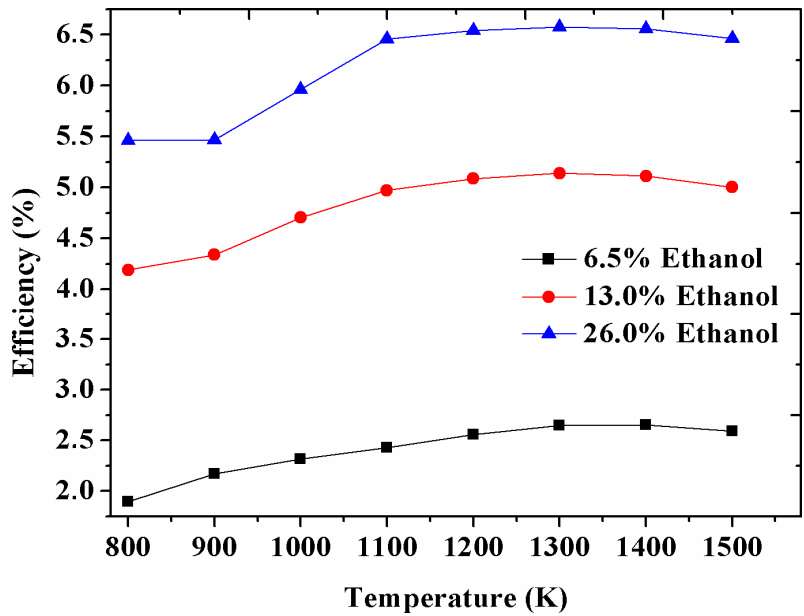

**Figure 10.** The efficiency of researched plasma chemical reactor for different mixtures

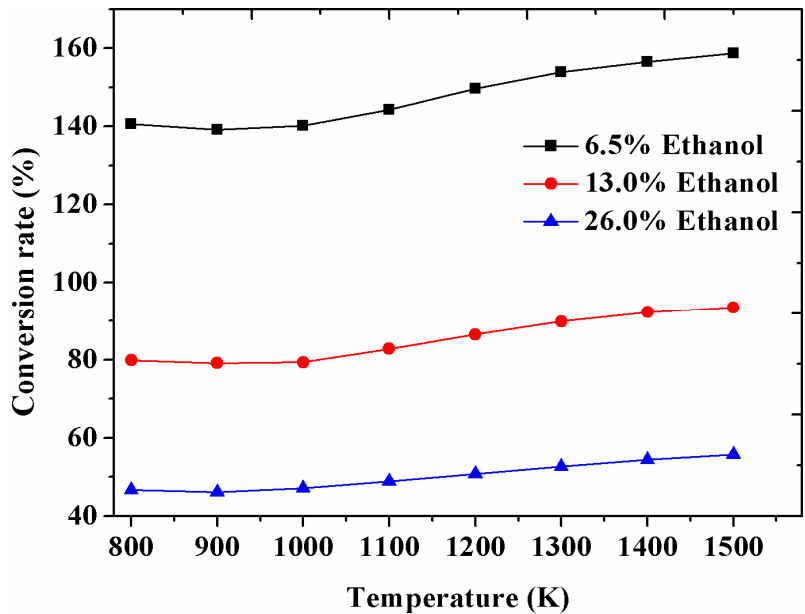

**Figure 11.** Conversion rate of researched plasma chemical reactor for different mixtures

## 5. Sensitivity analysis

Since hydrogen and carbon oxide are the main products at the outlet of the reactor, the sensitivity analysis was done specifically for these components. The value of each rate constant $k_j$ from table 2 was increased or decreased by factor of 10 and new [$H_2$] and [CO] were calculated for each change in $k_j$. Logarithmic sensitivity is defined by:

$$pS = \frac{log(Conc'/Conc)}{log(k'_j/k_j)}. \qquad (13)$$

Here $k_j$ is changed to a new value $k'_j$ and the concentration is changed from its old value *Conc* to a new value *Conc'*. Sensitivities were evaluated for each mixture at different temperatures. They could change significantly for different initial conditions. Figures 12-13 present the results of calculations for 6.5% concentration of ethanol in liquid and at $T = 1500$ K. For each process the lower bar is the value for increased rate constant and the upper bar is the value for the decreased one. One may notice that there are identical reactions for $H_2$ and CO. Additionally, the number of reactions for $H_2$ is much lower than for CO. All hydrogen reactions include $CH_3$ and H. The $CH_3$ radical is the main component in methane generation and the H atom participates in all reactions involving production of molecular hydrogen.

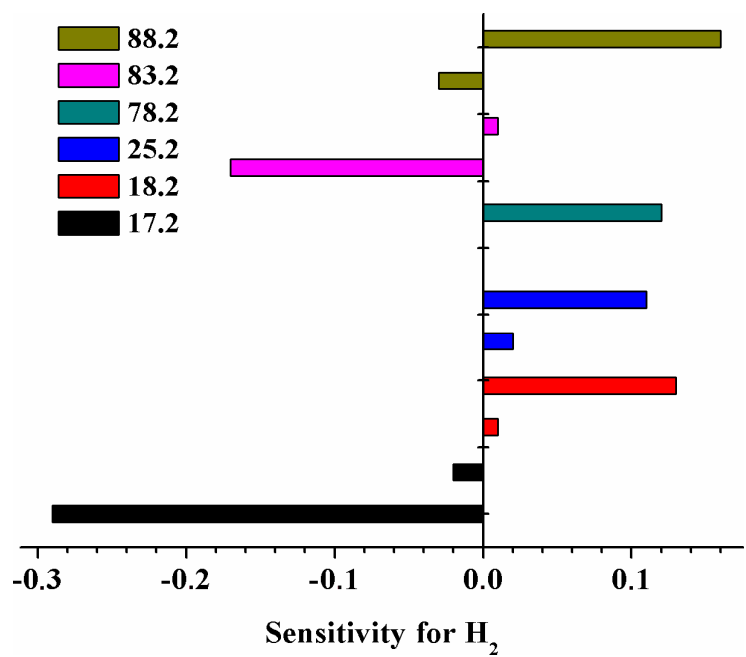

**Figure 12.** Sensitivity of [$H_2$] at 6.5% concentration of ethanol in the solution at 1500 K

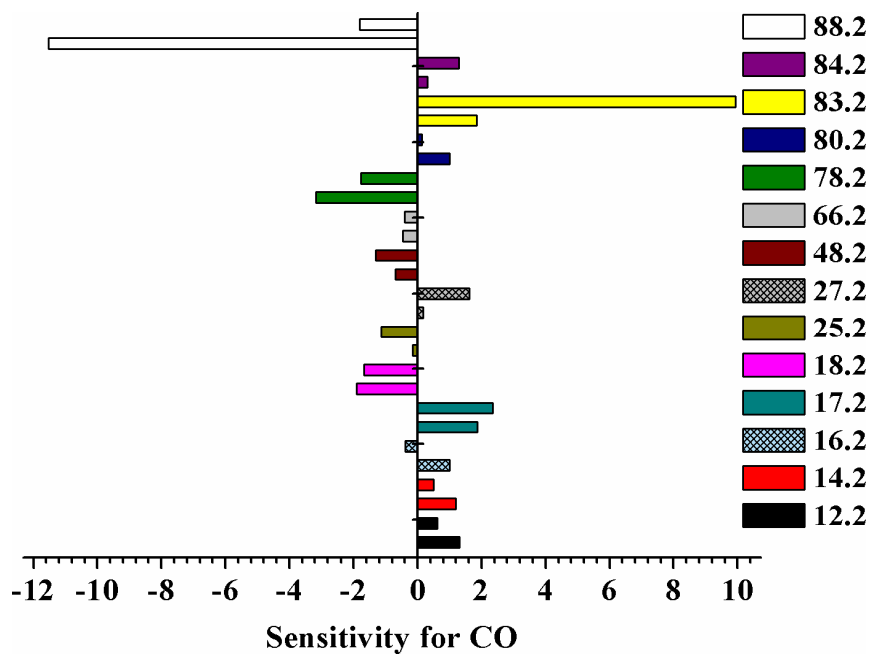

**Figure 13.** Sensitivity of [CO] at 6.5% concentration of ethanol in the solution at 1500 K

Figure 12 shows the reactions that have the most significant influence on [$H_2$]. One can notice, that the process (R4.2) is not among them. The temperature dependence of sensitivity of the molecular hydrogen concentration from rate constant of process (R4.2) is shown on figure 14. It is presented only by the curves for decreased rate constant. We don't present the dependence for increased $k$, since the behavior is similar. It is seen that the highest sensitivity is near 0.1 and it decreases when $T$ grows. It reaches near zero levels at high temperatures. Calculations indicate that the carbon oxide concentration has no sensitivity on the process (R4.2).

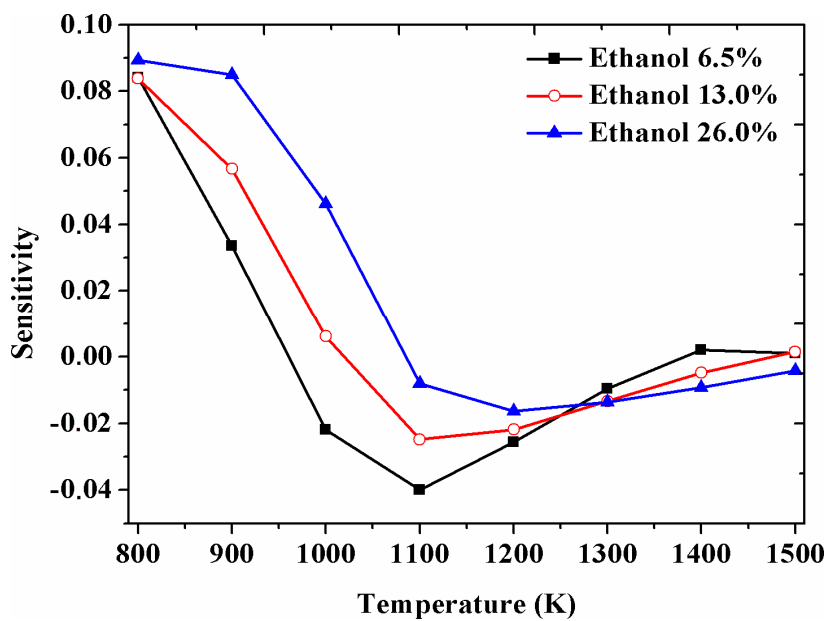

**Figure 14.** Sensitivity of [$H_2$] to rate constant of reaction (R4.2)

## 6. Conclusions

Kinetics in "tornado" type electrical discharge in ethanol/water/air mixture in the discharge and post-discharge regions were investigated. It was shown, that the ethanol conversion was taking place at both regions. The discharge region acts as a catalyst for the generation of active species (O, OH, H etc), gas heating and ethanol/water conversion into molecular hydrogen, carbon oxides and hydrocarbons $C_xH_y$. In the post-discharge region the conversion of ethanol continues, with the additional process of conversion of hydrocarbons $C_xH_y$.

Numerical simulation of kinetics showed, that the main channels of $H_2$ generation in plasma were ethanol abstraction for the first 10-100 ?s, and hydrocarbons abstraction afterwards. Additionally, the conditions when the reaction between $H_2O$ and hydrogen atoms was the main channel of $H_2$ production were found.

A kinetic mechanism, which used to describe adequately the chemistry of main components, was proposed. The model did not account for nitrogen-containing species, and nitrogen was considered only as a third body in recombination and dissociation reactions. The comparison between experiments and calculations showed, that the mechanism can describe adequately the concentrations of the main components ($H_2$, CO, $CO_2$, $CH_4$, $C_2H_4$, $C_2H_6$ and $C_2H_2$). Sensitivity analysis for the two main components ($H_2$ and CO) revealed the most important chemical reactions.

The highest hydrogen yield was reached when concentration of ethanol in the solution was 13%. However, the use of post-discharge region was more beneficial at 6.5% $C_2H_5OH$ concentration, because it increased [$H_2$] by 30% as compared to the discharge region. Additionally, this regime had the highest conversion rate among the investigated regimes. At the same time, the plasma chemical reactor had the highest efficiency when the concentration of ethanol in the solution was 26%.

**Acknowledgements**

This paper was supported by the Air Force Office of Scientific Research (USA). Project number is P354 "Plasma conversion of liquid hydrocarbon fuels in non-equilibrium plasma liquid systems". Authors thank Dr. Julian M. Tishkoff from the U.S. Air Force Office of Scientific Research and Mr. Daniel Shafer from the Israel Institute of Technology for helpful discussions and advice during the research.

**Table 1.** Reactions which were taken into account in the EEDF calculations

| № | Reaction | Reference |
|---|---|---|
| 1.1 | $H_2O + e \rightarrow H_2O((100)+(010)) + e$ | 14 |
| 2.1 | $H_2O + e \rightarrow H_2O(010) + e$ | 14 |
| 3.1 | $H_2O + e \rightarrow OH + H + e$ | 14 |
| 4.1 | $H_2O + e \rightarrow H_2O^+ + 2e$ | 14 |
| 5.1 | $H_2O + e \rightarrow H_2O(J = 0\text{-}0) + e$ | 14 |
| 6.1 | $H_2O + e \rightarrow H_2O(J = 0\text{-}1) + e$ | 14 |
| 7.1 | $H_2O + e \rightarrow H_2O(J = 0\text{-}2) + e$ | 14 |
| 8.1 | $H_2O + e \rightarrow H_2O(J = 0\text{-}3) + e$ | 14 |
| 9.1 | $N_2 + e \rightarrow N_2(A^3\Sigma_u^+) + e$ | 15 |
| 10.1 | $N_2 + e \rightarrow N_2(a^1\Pi_g) + e$ | 16 |
| 11.1 | $N_2 + e \rightarrow N_2(v) + e$ | 17 |
| 12.1 | $N_2 + e \rightarrow N + N + e$ | 18 |
| 13.1 | $N_2 + e \rightarrow N_2^+ + 2e$ | 19 |
| 14.1 | $O_2 + e \rightarrow O + O + e$ | 11 |
| 15.1 | $O_2 + e \rightarrow O_2(^1\Delta_g) + e$ | 20 |
| 16.1 | $O_2 + e \rightarrow O_2^+ + 2e$ | 19 |
| 17.1 | $C_2H_5OH + e \rightarrow CH_3 + CH_2OH + e$ | - |
| 18.1 | $C_2H_5OH + e \rightarrow C_2H_5 + OH + e$ | - |
| 19.1 | $C_2H_5OH + e \rightarrow CH_3CHOH + H + e$ | - |
| 20.1 | $C_2H_5OH + e \rightarrow C_2H_5OH^+ + 2e$ | 21 |

**Table 2.** Scheme of chemical reactions for ethanol conversion

| № | Reaction | Rate constant, cm$^3$/s or cm$^6$/s | Ref. |
|---|---|---|---|
| 1.2 | $C_2H_5OH+M \rightarrow CH_3+CH_2OH+M$ | $4.98 \cdot 10^{-6} \cdot \exp(-37997/T)$ | 29 |
| 2.2 | $C_2H_5OH+OH \rightarrow C_2H_5O+H_2O$ | $4.98 \cdot 10^{-11} \cdot \exp(-3000/T)$ | 29 |
| 3.2 | $C_2H_5OH+O \rightarrow C_2H_5O+OH$ | $3.1 \cdot 10^{-17} \cdot T^{1.85} \cdot \exp(-919.7/T)$ | 24 |
| 4.2 | $C_2H_5OH+H \rightarrow C_2H_5O+H_2$ | $5.41 \cdot 10^{-13} \cdot (T/298)^{2.53} \cdot \exp(-1722/T)$ | 30 |
| 5.2 | $C_2H_5OH+H \rightarrow C_2H_5+H_2O$ | $9.8 \cdot 10^{-13} \cdot \exp(-1740/T)$ | 31 |
| 6.2 | $C_2H_5OH+CH_3 \rightarrow C_2H_5O+CH_4$ | $1.22 \cdot 10^{-21} \cdot T^{2.99} \cdot \exp(-4000/T)$ | 24 |
| 7.2 | $C_2H_5OH+HO_2 \rightarrow C_2H_5O+H_2O_2$ | $4.2 \cdot 10^{-12} \cdot \exp(-12000/T)$ | 24 |
| 8.2 | $OH+OH+M \rightarrow H_2O_2+M$ | $6.05 \cdot 10^{-31} \cdot (T/298)^{-3}$ | 32 |
| 9.2 | $OH+OH \rightarrow H_2O+O$ | $5.6 \cdot 10^{-20} \cdot T^{2.42} \cdot \exp(973.1/T)$ | 28 |
| 10.2 | $O+OH \rightarrow O_2+H$ | $4.55 \cdot 10^{-12} \cdot (T/298)^{0.4} \cdot \exp(372/T)$ | 33 |
| 11.2 | $O+O+M \rightarrow O_2+M$ | $9.26 \cdot 10^{-34} \cdot (T/298)^{-1}$ | 2 |
| 12.2 | $H+OH+M \rightarrow H_2O+M$ | $6.86 \cdot 10^{-31} \cdot (T/298)^{-2}$ | 34 |
| 13.2 | $H+O+M \rightarrow OH+M$ | $4.36 \cdot 10^{-32} \cdot (T/298)^{-1}$ | 32 |
| 14.2 | $H+H+M \rightarrow H_2+M$ | $1.9 \cdot 10^{-30} \cdot (T/298)^{-1}$ | 24 |
| 15.2 | $CH_3+OH \rightarrow CH_2O+H_2$ | $2.59 \cdot 10^{-13} \cdot (T/298)^{-0.53} \cdot \exp(-5440/T)$ | 35 |
| 16.2 | $CH_3+O \rightarrow CH_2O+H$ | $1.4 \cdot 10^{-10}$ | 34 |
| 17.2 | $CH_3+H+M \rightarrow CH_4+M$ | $2.68 \cdot 10^{-28} \cdot (T/298)^{-2.98} \cdot \exp(-635/T)$ | 36 |
| 18.2 | $CH_3+CH_3+M \rightarrow C_2H_6+M$ | $1.68 \cdot 10^{-24} \cdot (T/298)^{-7} \cdot \exp(-1390/T)$ | 37 |
| 19.2 | $HO_2+OH \rightarrow H_2O+O_2$ | $4.8 \cdot 10^{-11} \cdot \exp(250/T)$ | 37 |
| 20.2 | $HO_2+O \rightarrow OH+O_2$ | $2.9 \cdot 10^{-11} \cdot \exp(200/T)$ | 32 |
| 21.2 | $HO_2+H \rightarrow OH+OH$ | $2.8 \cdot 10^{-10} \cdot \exp(-440/T)$ | 32 |
| 22.2 | $HO_2+H \rightarrow H_2+O_2$ | $1.1 \cdot 10^{-10} \cdot \exp(-1070/T)$ | 32 |
| 23.2 | $HO_2+CH_3 \rightarrow CH_3O+OH$ | $3.00 \cdot 10^{-11}$ | 32 |
| 24.2 | $HO_2+HO_2 \rightarrow H_2O_2+O_2$ | $3.00 \cdot 10^{-12}$ | 32 |
| 25.2 | $C_2H_5O+M \rightarrow CH_3+CH_2O+M$ | $1.7 \cdot 10^{-10} \cdot \exp(-12600/T)$ | 24 |
| 26.2 | $O_2+M \rightarrow O+O+M$ | $1.01 \cdot 10^{-8} \cdot (T/298)^{-1} \cdot \exp(-59429/T)$ | 32 |
| 27.2 | $O_2+H \rightarrow O+OH$ | $1.62 \cdot 10^{-10} \cdot \exp(-7474/T)$ | 37 |
| 28.2 | $O_2+H+M \rightarrow HO_2+M$ | $1.78 \cdot 10^{-32} \cdot (T/298)^{-0.8}$ | 37 |
| 29.2 | $O_2+CH_3 \rightarrow CH_3O+O$ | $2.2 \cdot 10^{-10} \cdot \exp(-15800/T)$ | 34 |
| 30.2 | $O_2+C_2H_5O \rightarrow CH_3CHO+HO_2$ | $1.66 \cdot 10^{-11} \cdot \exp(-2800/T)$ | 29 |
| 31.2 | $CH_3CHO \rightarrow CH_3+HCO$ | $2 \cdot 10^{15} \cdot \exp(-39811/T)$ | 2 |
| 32.2 | $CH_3CHO+OH \rightarrow H_2O+CH_3CO$ | $1.66 \cdot 10^{-11}$ | 2 |
| 33.2 | $CH_3CHO+O \rightarrow OH+CH_3CO$ | $8.3 \cdot 10^{-12} \cdot \exp(-902/T)$ | 2 |
| 34.2 | $CH_3CHO+H \rightarrow H_2+CH_3CO$ | $6.64 \cdot 10^{-11} \cdot \exp(-2117/T)$ | 2 |
| 35.2 | $CH_3CHO+CH_3 \rightarrow CH_4+CH_3CO$ | $2.97 \cdot 10^{-16} \cdot (T/298)^{5.64} \cdot \exp(-1240/T)$ | 34 |
| 36.2 | $CH_3CHO+O_2 \rightarrow HO_2+CH_3CO$ | $5 \cdot 10^{-11} \cdot \exp(-19700/T)$ | 34 |
| 37.2 | $CH_3CO+M \rightarrow CH_3+CO+M$ | $0.00683 \cdot (T/298)^{-8.62} \cdot \exp(-11284/T)$ | 32 |
| 38.2 | $CH_2OH+M \rightarrow CH_2O+H+M$ | $4.89 \cdot 10^{-5} \cdot (T/298)^{-2.5} \cdot \exp(-17205/T)$ | 38 |
| 39.2 | $CH_2OH+OH \rightarrow CH_2O+H_2O$ | $4.00 \cdot 10^{-11}$ | 38 |
| 40.2 | $CH_2OH+O \to CH_2O+OH$ | $7.00 \cdot 10^{-11}$ | 38 |
| 41.2 | $CH_2OH+H \to CH_2O+H_2$ | $10^{-11}$ | 38 |
| 42.2 | $CH_2OH+H \to CH_3+OH$ | $1.6 \cdot 10^{-10}$ | 38 |
| 43.2 | $CH_2OH+HO_2 \to CH_2O+H_2O_2$ | $2.00 \cdot 10^{-11}$ | 38 |
| 44.2 | $CH_2OH+O_2 \to CH_2O+HO_2$ | $2.00 \cdot 10^{-12}$ | 38 |
| 45.2 | $CH_2OH+CH_2OH \to CH_3OH+CH_2O$ | $8.00 \cdot 10^{-12}$ | 38 |
| 46.2 | $CH_4+OH \to CH_3+H_2O$ | $8.77 \cdot 10^{-13} \cdot (T/298)^{1.83} \cdot \exp(-1400/T)$ | 34 |
| 47.2 | $CH_4+O \to CH_3+OH$ | $8.32 \cdot 10^{-12} \cdot (T/298)^{1.56} \cdot \exp(-4270/T)$ | 34 |
| 48.2 | $CH_4+H \to CH_3+H_2$ | $5.82 \cdot 10^{-13} \cdot (T/298)^{3} \cdot \exp(-4045/T)$ | 34 |
| 49.2 | $CH_4+HO_2 \to CH_3+H_2O_2$ | $3 \cdot 10^{-13} \cdot \exp(-9350/T)$ | 32 |
| 50.2 | $CH_4+O_2 \to CH_3+HO_2$ | $6.7 \cdot 10^{-11} \cdot \exp(-28640/T)$ | 32 |
| 51.2 | $CH_2O+M \to CO+H_2+M$ | $3.49 \cdot 10^{-9} \cdot \exp(-17620/T)$ | 39 |
| 52.2 | $CH_2O+OH \to HCO+H_2O$ | $4.73 \cdot 10^{-12} \cdot (T/298)^{1.18} \cdot \exp(225/T)$ | 34 |
| 53.2 | $CH_2O+O \to HCO+OH$ | $1.78 \cdot 10^{-11} \cdot (T/298)^{0.57} \cdot \exp(-1390/T)$ | 34 |
| 54.2 | $CH_2O+H \to HCO+H_2$ | $1.5 \cdot 10^{-11} \cdot (T/298)^{1.05} \cdot \exp(-1650/T)$ | 34 |
| 55.2 | $CH_2O+CH_3 \to CH_4+HCO$ | $1.6 \cdot 10^{-16} \cdot (T/298)^{6.1} \cdot \exp(-990/T)$ | 37 |
| 56.2 | $CH_2O+HO_2 \to HCO+H_2O_2$ | $3.3 \cdot 10^{-12} \cdot \exp(-5870/T)$ | 32 |
| 57.2 | $CH_3O+M \to CH_2O+H+M$ | $9 \cdot 10^{-11} \cdot \exp(-6790/T)$ | 37 |
| 58.2 | $CH_3O+OH \to CH_2O+H_2O$ | $3.00 \cdot 10^{-11}$ | 32 |
| 59.2 | $CH_3O+O \to CH_2O+OH$ | $1.00 \cdot 10^{-11}$ | 32 |
| 60.2 | $CH_3O+H \to CH_2O+H_2$ | $3.30 \cdot 10^{-11}$ | 32 |
| 61.2 | $CH_3O+H \to CH_3+OH$ | $5.30 \cdot 10^{-11}$ | 25 |
| 62.2 | $CH_3O+HO_2 \to CH_2O+H_2O_2$ | $5.00 \cdot 10^{-13}$ | 32 |
| 63.2 | $CH_3O+O_2 \to CH_2O+HO_2$ | $1.1 \cdot 10^{-13} \cdot \exp(-1310/T)$ | 32 |
| 64.2 | $CH_3O+CH_2OH \to CH_3OH+CH_2O$ | $4.00 \cdot 10^{-11}$ | 38 |
| 65.2 | $CH_3O+CH_3O \to CH_3OH+CH_2O$ | $10^{-10}$ | 32 |
| 66.2 | $HCO+M \to CO+H+M$ | $2.6 \cdot 10^{-10} \cdot \exp(-7930/T)$ | 37 |
| 67.2 | $HCO+OH \to CO+H_2O$ | $1.70 \cdot 10^{-10}$ | 34 |
| 68.2 | $HCO+O \to CO+OH$ | $5.00 \cdot 10^{-11}$ | 34 |
| 69.2 | $HCO+H \to CO+H_2$ | $1.50 \cdot 10^{-10}$ | 34 |
| 70.2 | $HCO+CH_3 \to CH_4+CO$ | $2.00 \cdot 10^{-10}$ | 32 |
| 71.2 | $HCO+HO_2 \to CO_2+OH+H$ | $5.00 \cdot 10^{-11}$ | 32 |
| 72.2 | $HCO+O_2 \to CO+HO_2$ | $8.5 \cdot 10^{-11} \cdot \exp(-850/T)$ | 32 |
| 73.2 | $HCO+CH_2OH \to CH_3OH+CO$ | $2.00 \cdot 10^{-10}$ | 38 |
| 74.2 | $HCO+CH_2OH \to CH_2O+CH_2O$ | $3.00 \cdot 10^{-10}$ | 38 |
| 75.2 | $HCO+CH_3O \to CH_3OH+CO$ | $1.50 \cdot 10^{-10}$ | 32 |
| 76.2 | $HCO+HCO \to CH_2O+CO$ | $3.00 \cdot 10^{-11}$ | 32 |
| 77.2 | $HCO+HCO \to H_2+CO+CO$ | $5.00 \cdot 10^{-12}$ | 25 |
| 78.2 | $CO+OH \to CO_2+H$ | $5.4 \cdot 10^{-14} \cdot (T/298)^{1.5} \cdot \exp(250/T)$ | 34 |
| 79.2 | $CO+O+M \to CO_2+M$ | $1.46 \cdot 10^{-34} \cdot \exp(2285/T)$ | 2 |
| 80.2 | $CO+H+M \to HCO+M$ | $5.3 \cdot 10^{-34} \cdot \exp(-370/T)$ | 37 |
| 81.2 | $CO+HO_2 \to CO_2+OH$ | $2.5 \cdot 10^{-10} \cdot \exp(-11900/T)$ | 32 |
| 82.2 | $CO+CH_3O \to CH_3+CO_2$ | $2.6 \cdot 10^{-11} \cdot \exp(-5940/T)$ | 32 |
| 83.2 | $H_2+OH \to H_2O+H$ | $1.55 \cdot 10^{-12} \cdot (T/298)^{1.6} \cdot \exp(-1660/T)$ | 34 |
| 84.2 | $H_2+O \to H+OH$ | $3.43 \cdot 10^{-13} \cdot (T/298)^{2.67} \cdot \exp(-3160/T)$ | 34 |
| 85.2 | $H_2+O_2 \to OH+OH$ | $2.8 \cdot 10^{-11} \cdot \exp(-24100/T)$ | 24 |
| 86.2 | $H_2+O_2 \to H+HO_2$ | $2.4 \cdot 10^{-10} \cdot \exp(-28500/T)$ | 32 |
| 87.2 | $H_2O+O \to OH+OH$ | $1.25 \cdot 10^{-11} \cdot (T/298)^{1.3} \cdot \exp(-8605/T)$ | 32 |
| 88.2 | $H_2O+H \to H_2+OH$ | $5.17 \cdot 10^{-12} \cdot (T/298)^{1.9} \cdot \exp(-9265/T)$ | 32 |
| 89.2 | $H_2O_2+M \to OH+OH+M$ | $3 \cdot 10^{-8} \cdot \exp(-21600/T)$ | 34 |
| 90.2 | $H_2O_2+OH \to H_2O+HO_2$ | $2.9 \cdot 10^{-12} \cdot \exp(-160/T)$ | 32 |
| 91.2 | $H_2O_2+O \to OH+HO_2$ | $1.42 \cdot 10^{-12} \cdot (T/298)^2 \cdot \exp(-2000/T)$ | 32 |
| 92.2 | $H_2O_2+H \to HO_2+H_2$ | $8 \cdot 10^{-11} \cdot \exp(-4000/T)$ | 32 |
| 93.2 | $H_2O_2+H \to H_2O+OH$ | $4 \cdot 10^{-11} \cdot \exp(-2000/T)$ | 32 |
| 94.2 | $H_2O_2+O_2 \to HO_2+HO_2$ | $9 \cdot 10^{-11} \cdot \exp(-20000/T)$ | 32 |
| 95.2 | $C_2H_6 \to CH_3+CH_3$ | $1.54 \cdot 10^{18} \cdot (T/298)^{-1.24} \cdot \exp(-45700/T)$ | 37 |
| 96.2 | $C_2H_6+OH \to C_2H_5+H_2O$ | $1.06 \cdot 10^{-12} \cdot (T/298)^2 \cdot \exp(-435/T)$ | 34 |
| 97.2 | $C_2H_6+O \to C_2H_5+OH$ | $6.1 \cdot 10^{-11} \cdot (T/298)^{0.6} \cdot \exp(-3680/T)$ | 32 |
| 98.2 | $C_2H_6+H \to C_2H_5+H_2$ | $4.2 \cdot 10^{-13} \cdot (T/298)^{3.5} \cdot \exp(-2600/T)$ | 32 |
| 99.2 | $C_2H_6+CH_3 \to CH_4+C_2H_5$ | $7.18 \cdot 10^{-15} \cdot (T/298)^4 \cdot \exp(-4169/T)$ | 32 |
| 100.2 | $C_2H_5 \to C_2H_4+H$ | $4.31 \cdot 10^{12} \cdot (T/298)^{1.19} \cdot \exp(-18722/T)$ | 32 |
| 101.2 | $C_2H_5+O_2 \to C_2H_4+HO_2$ | $1.4 \cdot 10^{-12} \cdot \exp(-1950/T)$ | 32 |
| 102.2 | $C_2H_4+OH \to C_2H_3+H_2O$ | $1.66 \cdot 10^{-13} \cdot (T/298)^{2.75} \cdot \exp(-2100/T)$ | 32 |
| 103.2 | $C_2H_4+O \to CH_3+HCO$ | $1.5 \cdot 10^{-12} \cdot (T/298)^{1.55} \cdot \exp(-215/T)$ | 32 |
| 104.2 | $C_2H_4+H \to C_2H_3+H_2$ | $4 \cdot 10^{-12} \cdot (T/298)^{2.53} \cdot \exp(-6160/T)$ | 32 |
| 105.2 | $C_2H_3+M \to C_2H_2+H+M$ | $0.192 \cdot (T/298)^{-7.5} \cdot \exp(-22900/T)$ | 34 |
| 106.2 | $C_2H_3+ C_2H_5 \to C_2H_4+C_2H_4$ | $2.70 \cdot 10^{-11}$ | 24 |
| 107.2 | $C_2H_2+OH \to C_2H+H_2O$ | $1.03 \cdot 10^{-13} \cdot (T/298)^{2.68} \cdot \exp(-6060/T)$ | 32 |
| 108.2 | $C_2H_2+OH \to CH_3+CO$ | $6.34 \cdot 10^{-18} \cdot (T/298)^4 \cdot \exp(1006/T)$ | 40 |
| 109.2 | $C_2H_2+O \to CH_2+CO$ | $3.5 \cdot 10^{-12} \cdot (T/298)^{1.5} \cdot \exp(-854/T)$ | 2 |
| 110.2 | $C_2H_2+H \to C_2H+H_2$ | $10^{-10} \cdot \exp(-11200/T)$ | 32 |
| 111.2 | $C_2H+O \to CO+CH$ | $1.70 \cdot 10^{-11}$ | 34 |
| 112.2 | $C_2H+O_2 \to HCO+CO$ | $4.00 \cdot 10^{-12}$ | 32 |
| 113.2 | $CH_2+OH \to CH+H_2O$ | $1.9 \cdot 10^{-17} \cdot T^{2.0} \cdot \exp(-1510/T)$ | 24 |
| 114.2 | $CH_2+O \to CH+OH$ | $4.98 \cdot 10^{-10} \cdot \exp(-6000/T)$ | 41 |
| 115.2 | $CH_2+H \to CH+H_2$ | $10^{-11} \cdot \exp(900/T)$ | 34 |
| 116.2 | $CH_2+O_2 \to HCO+OH$ | $2.2 \cdot 10^{-4} \cdot (T/298)^{-3.3} \cdot \exp(-143.2/T)$ | 24 |
| 117.2 | $CH+O_2 \to HCO+O$ | $5.50 \cdot 10^{-11}$ | 24 |
| 118.2 | $CH+O_2 \to CO+OH$ | $3.30 \cdot 10^{-11}$ | 42 |
| 119.2 | $CH_3OH+M \to CH_3+OH+M$ | $3.32 \cdot 10^{-7} \cdot \exp(-34399/T)$ | 2 |
| 120.2 | $CH_3OH+M \to CH_2OH+H+M$ | $61.4 \cdot (T/298)^{-8} \cdot \exp(-45295/T)$ | 43 |
| 121.2 | $CH_3OH+OH \to CH_2OH+H_2O$ | $2.12 \cdot 10^{-13} \cdot (T/298)^2 \cdot \exp(423/T)$ | 44 |
| 122.2 | $CH_3OH+OH \rightarrow CH_3O+H_2O$ | $1.66 \cdot 10^{-11} \cdot \exp(-854/T)$ | 2 |
| 123.2 | $CH_3OH+O \rightarrow CH_2OH+OH$ | $5.7 \cdot 10^{-11} \cdot \exp(-2750/T)$ | 45 |
| 124.2 | $CH_3OH+H \rightarrow CH_2OH+H_2$ | $2.42 \cdot 10^{-12} \cdot (T/298)^2 \cdot \exp(-2273/T)$ | 44 |
| 125.2 | $CH_3OH+H \rightarrow CH_3O+H_2$ | $6.64 \cdot 10^{-11} \cdot \exp(-3067/T)$ | 2 |
| 126.2 | $CH_3OH+CH_3 \rightarrow CH_2OH+CH_4$ | $4.38 \cdot 10^{-15} \cdot (T/298)^{3.2} \cdot \exp(-3609/T)$ | 38 |
| 127.2 | $CH_3OH+HO_2 \rightarrow CH_2OH+H_2O_2$ | $1.6 \cdot 10^{-13} \cdot \exp(-6330/T)$ | 38 |
| 128.2 | $CH_3OH+O_2 \rightarrow CH_2OH+HO_2$ | $3.4 \cdot 10^{-11} \cdot \exp(-22600/T)$ | 38 |
| 129.2 | $CH_3OH+CH_3O \rightarrow CH_3OH+CH_2OH$ | $5 \cdot 10^{-13} \cdot \exp(-2050/T)$ | 38 |
| 130.2 | $CH_3OH+HCO \rightarrow CH_2OH+CH_2O$ | $2.4 \cdot 10^{-13} \cdot (T/298)^{2.9} \cdot \exp(-6596/T)$ | 38 |

**Table 3.** Electron-molecular reactions which were taken into account in the plasma kinetics investigation

| № | Reaction | Threshold, eV | $W_{ei}$, cm$^3$/s | | | $S_{ei}$, s$^{-1}$ | Ref. |
|---|---|---|---|---|---|---|---|
| | | | 800 K | 1200 K | 1500 K | | |
| 1.3 | $C_2H_5OH+e \to C_2H_5+OH+e$ | 7.90 | $3.08 \cdot 10^{-9}$ | $6.46 \cdot 10^{-9}$ | $8.83 \cdot 10^{-9}$ | 203.74 | * |
| 2.3 | $C_2H_5OH+e \to C_2H_5O+H+e$ | 7.82 | $3.14 \cdot 10^{-9}$ | $6.55 \cdot 10^{-9}$ | $8.93 \cdot 10^{-9}$ | 207.88 | * |
| 3.3 | $C_2H_5OH+e \to CH_2OH+CH_3+e$ | 7.38 | $3.50 \cdot 10^{-9}$ | $7.05 \cdot 10^{-9}$ | $9.50 \cdot 10^{-9}$ | 231.76 | * |
| 4.3 | $OH+e \to O+H+e$ | 8.80 | $2.44 \cdot 10^{-9}$ | $5.51 \cdot 10^{-9}$ | $7.75 \cdot 10^{-9}$ | 161.37 | * |
| 5.3 | $CH_3+e \to CH_2+H+e$ | 9.50 | $2.02 \cdot 10^{-9}$ | $4.84 \cdot 10^{-9}$ | $6.97 \cdot 10^{-9}$ | 133.49 | * |
| 6.3 | $HO_2+e \to OH+O+e$ | 5.60 | $5.26 \cdot 10^{-9}$ | $9.32 \cdot 10^{-9}$ | $1.19 \cdot 10^{-8}$ | 347.62 | * |
| 7.3 | $C_2H_5O+e \to C_2H_4+OH+e$ | 3.46 | $7.90 \cdot 10^{-9}$ | $1.24 \cdot 10^{-8}$ | $1.51 \cdot 10^{-8}$ | 522.38 | * |
| 8.3 | $C_2H_5O+e \to CH_3CHO+H+e$ | 7.56 | $3.35 \cdot 10^{-9}$ | $6.84 \cdot 10^{-9}$ | $9.26 \cdot 10^{-9}$ | 221.76 | * |
| 9.3 | $C_2H_5O+e \to CH_3+CH_2O+e$ | 5.12 | $5.80 \cdot 10^{-9}$ | $9.98 \cdot 10^{-9}$ | $1.26 \cdot 10^{-8}$ | 383.89 | * |
| 10.3 | $O_2+e \to O+O+e$ | 6.00 | $4.82 \cdot 10^{-9}$ | $8.78 \cdot 10^{-9}$ | $1.14 \cdot 10^{-8}$ | 318.95 | 11 |
| 11.3 | $CH_3CHO+e \to CH_3+HCO+e$ | 7.04 | $3.80 \cdot 10^{-9}$ | $7.46 \cdot 10^{-9}$ | $9.94 \cdot 10^{-9}$ | 251.51 | * |
| 12.3 | $CH_3CHO+e \to C_2H_4+O+e$ | 9.72 | $1.90 \cdot 10^{-9}$ | $4.64 \cdot 10^{-9}$ | $6.73 \cdot 10^{-9}$ | 125.58 | * |
| 13.3 | $CH_3CHO+e \to CH_3CO+H+e$ | 7.60 | $3.32 \cdot 10^{-9}$ | $6.80 \cdot 10^{-9}$ | $9.21 \cdot 10^{-9}$ | 219.58 | * |
| 14.3 | $CH_3CO+e \to CH_3+CO+e$ | 1.04 | $1.12 \cdot 10^{-8}$ | $1.60 \cdot 10^{-9}$ | $1.87 \cdot 10^{-8}$ | 743.55 | * |
| 15.3 | $CH_3CO+e \to C_2H_3+O+e$ | 11.30 | $1.21 \cdot 10^{-9}$ | $3.41 \cdot 10^{-9}$ | $5.22 \cdot 10^{-9}$ | 79.93 | * |
| 16.3 | $CH_2OH+e \to CH_2+OH+e$ | 9.80 | $1.86 \cdot 10^{-9}$ | $4.57 \cdot 10^{-9}$ | $6.65 \cdot 10^{-9}$ | 122.80 | * |
| 17.3 | $CH_2OH+e \to CH_2O+H+e$ | 3.18 | $8.27 \cdot 10^{-9}$ | $1.28 \cdot 10^{-8}$ | $1.55 \cdot 10^{-8}$ | 547.26 | * |
| 18.3 | $CH_4+e \to CH_3+H+e$ | 4.50 | $2.29 \cdot 10^{-9}$ | $5.32 \cdot 10^{-9}$ | $7.60 \cdot 10^{-9}$ | 151.57 | 47 |
| 19.3 | $CH_2O+e \to CH_2+O+e$ | 15.10 | $3.05 \cdot 10^{-10}$ | $1.33 \cdot 10^{-9}$ | $2.38 \cdot 10^{-9}$ | 20.18 | * |
| 20.3 | $CH_2O+e \to HCO+H+e$ | 7.56 | $3.35 \cdot 10^{-9}$ | $6.84 \cdot 10^{-9}$ | $9.26 \cdot 10^{-9}$ | 221.76 | * |
| 21.3 | $CH_3O+e \to CH_3+O+e$ | 7.95 | $3.04 \cdot 10^{-9}$ | $6.40 \cdot 10^{-9}$ | $8.77 \cdot 10^{-9}$ | 201.18 | * |
| 22.3 | $CH_3O+e \to CH_2O+H+e$ | 7.56 | $3.35 \cdot 10^{-9}$ | $6.84 \cdot 10^{-9}$ | $9.26 \cdot 10^{-9}$ | 221.76 | * |
| 23.3 | $HCO+e \to CH+O+e$ | 16.90 | $1.86 \cdot 10^{-10}$ | $9.41 \cdot 10^{-10}$ | $1.78 \cdot 10^{-9}$ | 12.32 | * |
| 24.3 | $HCO+e \to CO+H+e$ | 1.60 | $1.05 \cdot 10^{-8}$ | $1.52 \cdot 10^{-8}$ | $1.80 \cdot 10^{-8}$ | 697.12 | * |
| 25.3 | $CO_2+e \to CO+O+e$ | 5.52 | $1.88 \cdot 10^{-10}$ | $4.37 \cdot 10^{-10}$ | $6.24 \cdot 10^{-10}$ | 12.41 | 47 |
| 26.3 | $H_2+e \to H+H+e$ | 7.80 | $1.67 \cdot 10^{-9}$ | $3.15 \cdot 10^{-9}$ | $4.06 \cdot 10^{-9}$ | 110.16 | 11 |
| 27.3 | $H_2O+e \to OH+H+e$ | 7.00 | $1.93 \cdot 10^{-9}$ | $3.80 \cdot 10^{-9}$ | $5.08 \cdot 10^{-9}$ | 127.43 | 14 |
| 28.3 | $H_2O_2+e \to HO_2+H+e$ | 7.56 | $3.35 \cdot 10^{-9}$ | $6.84 \cdot 10^{-9}$ | $9.26 \cdot 10^{-9}$ | 221.76 | * |
| 29.3 | $H_2O_2+e \to OH+OH+e$ | 4.44 | $6.63 \cdot 10^{-9}$ | $1.10 \cdot 10^{-8}$ | $1.36 \cdot 10^{-8}$ | 438.43 | * |
| 30.3 | $C_2H_6+e \to CH_3+CH_3+e$ | 7.66 | $3.27 \cdot 10^{-9}$ | $6.73 \cdot 10^{-9}$ | $9.14 \cdot 10^{-9}$ | 216.34 | * |
| 31.3 | $C_2H_6+e \to C_2H_5+H+e$ | 8.51 | $2.63 \cdot 10^{-9}$ | $5.80 \cdot 10^{-9}$ | $8.09 \cdot 10^{-9}$ | 174.19 | * |
| 32.3 | $C_2H_5+e \to C_2H_4+H+e$ | 3.38 | $8.00 \cdot 10^{-9}$ | $1.25 \cdot 10^{-8}$ | $1.52 \cdot 10^{-8}$ | 529.46 | * |
| 33.3 | $C_2H_5+e \to CH_2+CH_3+e$ | 8.64 | $2.55 \cdot 10^{-9}$ | $5.67 \cdot 10^{-9}$ | $7.94 \cdot 10^{-9}$ | 168.35 | * |
| 34.3 | $C_2H_4+e \to CH_2+CH_2+e$ | 5.12 | $5.80 \cdot 10^{-9}$ | $9.98 \cdot 10^{-9}$ | $1.26 \cdot 10^{-8}$ | 383.89 | 47 |
| 35.3 | $C_2H_4+e \to C_2H_3+H+e$ | 4.60 | $1.63 \cdot 10^{-12}$ | $4.20 \cdot 10^{-12}$ | $6.32 \cdot 10^{-12}$ | 0.11 | 47 |
| 36.3 | $C_2H_3+e \to C_2H_2+H+e$ | 3.48 | $7.87 \cdot 10^{-9}$ | $1.24 \cdot 10^{-8}$ | $1.51 \cdot 10^{-8}$ | 520.62 | * |
| 37.3 | $C_2H_3+e \to CH_2+CH+e$ | 5.12 | $5.80 \cdot 10^{-9}$ | $9.98 \cdot 10^{-9}$ | $1.26 \cdot 10^{-8}$ | 383.89 | * |

| 38.3 | $C_2H_2+e \rightarrow CH+CH+e$ | 5.12 | $5.80 \cdot 10^{-9}$ | $9.98 \cdot 10^{-9}$ | $1.26 \cdot 10^{-8}$ | 383.89 | * |
| 39.3 | $C_2H_2+e \rightarrow C_2H+H+e$ | 10.30 | $1.61 \cdot 10^{-9}$ | $4.15 \cdot 10^{-9}$ | $6.14 \cdot 10^{-9}$ | 106.57 | * |
| 40.3 | $CH_2+e \rightarrow CH+H+e$ | 8.92 | $2.36 \cdot 10^{-9}$ | $5.39 \cdot 10^{-9}$ | $7.62 \cdot 10^{-9}$ | 156.31 | * |
| 41.3 | $CH_3OH+e \rightarrow CH_3+OH+e$ | 7.94 | $3.05 \cdot 10^{-9}$ | $6.41 \cdot 10^{-9}$ | $8.78 \cdot 10^{-9}$ | 201.69 | * |
| 42.3 | $CH_3OH+e \rightarrow CH_2OH+H+e$ | 8.28 | $2.80 \cdot 10^{-9}$ | $6.05 \cdot 10^{-9}$ | $8.37 \cdot 10^{-9}$ | 184.92 | * |
| 43.3 | $CH_3OH+e \rightarrow CH_3O+H+e$ | 8.88 | $2.39 \cdot 10^{-9}$ | $5.43 \cdot 10^{-9}$ | $7.66 \cdot 10^{-9}$ | 157.97 | * |